\documentclass[sigconf]{acmart}

\AtBeginDocument{%
  \providecommand\BibTeX{{%
    \normalfont B\kern-0.5em{\scshape i\kern-0.25em b}\kern-0.8em\TeX}}}

\copyrightyear{2021}
\acmYear{2021}
\acmConference[CIKM '21]{Proceedings of the 29th ACM International Conference on Information \& Knowledge Management}{November 1-5, 2021}{Virtual}
\acmBooktitle{Proceedings of the 29th ACM International Conference on Information \& Knowledge Management (CIKM '21), November 1-5, 2021, Virtual}
\acmPrice{}

\usepackage{xspace}
\usepackage{xcolor}
\usepackage{enumitem}
\usepackage{bm}
\usepackage{graphicx}
\usepackage{subcaption}
\usepackage{amsmath}
\usepackage{url}
\usepackage{array,multirow}
\usepackage[utf8]{inputenc}
\usepackage[english]{babel}

\usepackage{amsthm}
\theoremstyle{plain}

\theoremstyle{remark}

\usepackage{adjustbox}
\usepackage[ruled,vlined]{algorithm2e} 

\newcommand{\modelname}{\texttt{SMART}\xspace}
\newcommand{\modelfullname}{
\underline{S}ocial \underline{M}edia enh\underline{A}nced pandemic su\underline{R}veillance \underline{T}echnique\xspace}

\begin{document}

\title{\#StayHome or \#Marathon? Social Media Enhanced Pandemic Surveillance on Spatial-temporal Dynamic Graphs}


\author{Yichao Zhou, Jyun-yu Jiang, Xiusi Chen, and Wei Wang}
\affiliation{
  \institution{Department of Computer Science, University of California, Los Angeles}
  \city{Los Angeles}
  \state{CA}
  \country{USA}
}
\email{{yz, jyunyu, xchen, weiwang}@cs.ucla.edu}



\begin{abstract}
COVID-19 has caused lasting damage to almost every domain in public health, society, and economy. To monitor the pandemic trend, existing studies rely on the aggregation of traditional statistical models and epidemic spread theory. In other words, historical statistics of COVID-19, as well as the population mobility data, become the essential knowledge for monitoring the pandemic trend. However, these solutions can barely provide precise prediction and satisfactory explanations on the long-term disease surveillance while the ubiquitous social media resources can be the key enabler for solving this problem. 
For example, serious discussions may occur on social media before and after some breaking events take place. These events, such as \textit{marathon} and \textit{parade}, may impact the spread of the virus. To take advantage of the social media data, we propose a novel framework, \modelfullname (\modelname), which is composed of two modules: (i) information extraction module to construct heterogeneous knowledge graphs based on the extracted events and relationships among them; (ii) time series prediction module to provide both short-term and long-term forecasts of the confirmed cases and fatality at the state-level in the United States and to discover risk factors for COVID-19 interventions. Extensive experiments show that our method largely outperforms the state-of-the-art baselines by 7.3\% and 7.4\% in confirmed case/fatality prediction, respectively. 
\end{abstract}




\maketitle

\section{Introduction}

\begin{figure}[t]
    \centering
    \includegraphics[width=\linewidth]{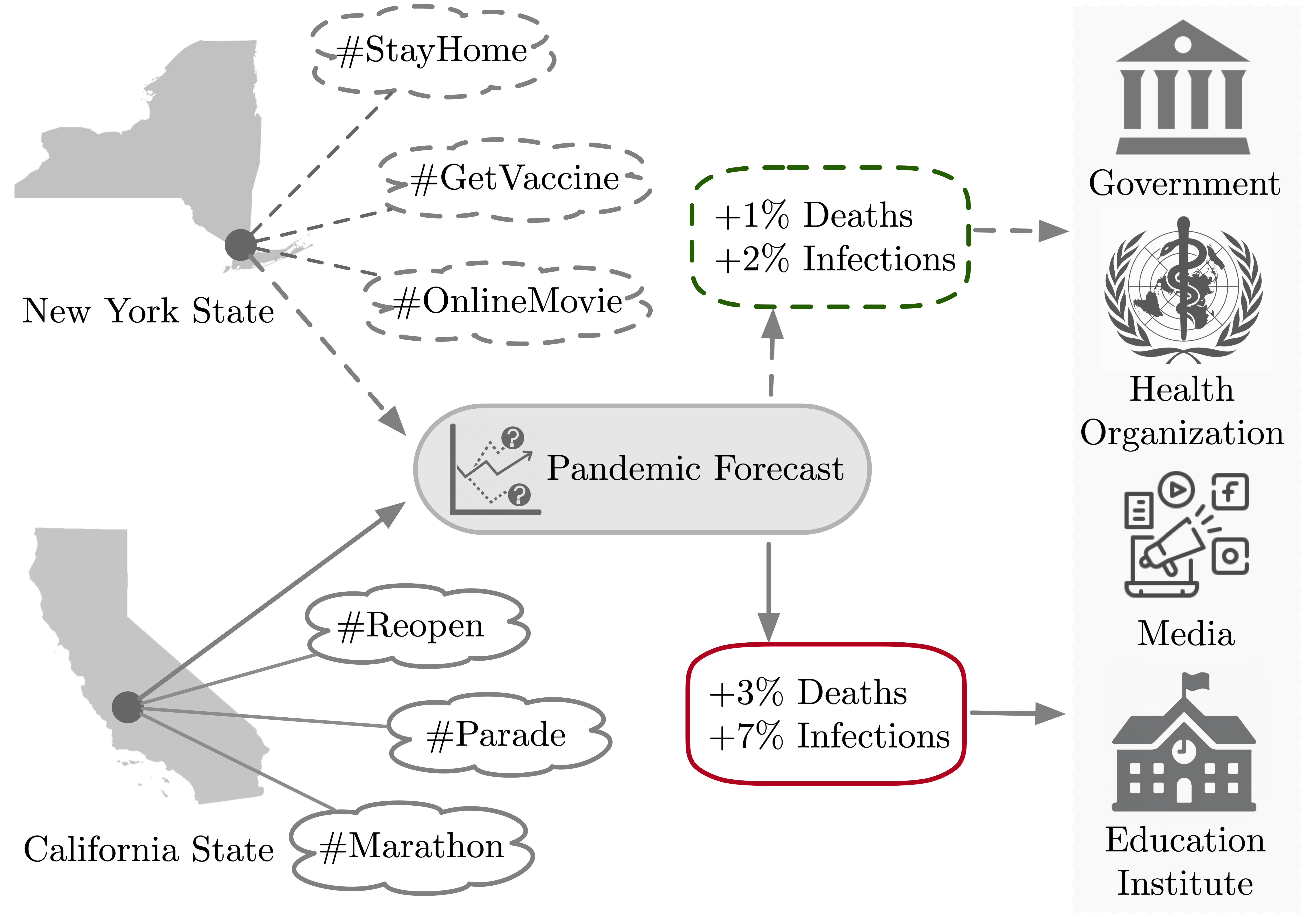}
    \caption{Social media users can serve as a ``social sensor'' for monitoring the pandemic trend. For example, some time-wise and location-wise prevailing entities in social media data such as ``Reopen'' and ``Parade'' indicate that people are likely to go out, leading to an increasing trend of virus transmission. The real-time forecasts will be delivered to the government, health organizations, all kinds of media, and education institutes for making intervention strategies.}
    \label{fig:intro}
\end{figure}

Over 200 countries and territories have been deeply impacted by the outbreak of the coronavirus disease 2019 (COVID-19). As of 2021 May, a total of 164 million cases and 3.4 million deaths were reported all over the world\footnote{\url{https://covid19.who.int/}}. It is critical to forecast the short-term and long-term trends of the epidemic, to help governments and health organizations determine the prevention strategies and help researchers understand the transmission characteristics of the virus.

Modeling the COVID-19 pandemic is challenging. Previous studies present three types of disease transmission approaches to explain and model the pandemic, which are exponential growth models~\cite{livadiotis2020statistical}, self-exiting branching process~\cite{kong2021evently}, and compartment models 
(e.g., Susceptible-infected-resistant (SIR)~\cite{kermack1927contribution},  Susceptible-Exposed-Infected-Removed (SEIR)~\cite{aron1984seasonality} and Herd Immunity~\cite{fine2011herd}). 
However, exponential growth models can only address the initial outbreak while self-exiting-branching process and compartment models favor the development and peak stages~\cite{bertozzi2020challenges}.
Besides, the pandemic trend varies dramatically across different locations and times in response to real-time breaking events. 
To tackle these challenges, some data-driven approaches~\cite{covid2020forecasting,altieri2020curating} that ensembles statistical and machine learning models emerge for monitoring the confirmed cases, fatality, and hospitalizations. \cite{panagopoulos2020transfer,gao2021stan} leverage graph neural networks to incorporate the population mobility data, i.e., how many people traveled from one place to another, to encode the underlying diffusion patterns into the learning process. 
However, these models take into consideration only a small number of homogeneous features. They are incapable of capturing potential risk factors and identifying various intervention mechanisms of this new pandemic as well. 

As the quarantine life takes over the world and people turn to online platforms for communication and information, social media become more influential than ever~\cite{han2020using,nabity2020inside}.
The vast collections of social media streams can capture local activities (e.g., public gatherings and vaccination progress) that may affect the transmission of the virus in real-time. Over 170 million tweets are posted every day in the United States related to observations, behaviors, and thoughts of individual users~\cite{clement_2020}. The social media users can be naturally treated as robust ``social sensors''~\cite{jiang2019enhancing} to unveil the surveillance evidence over time and space. For example, in Figure~\ref{fig:intro}, the severe discussions related to the coming social events such as ``Marathon'' and ``Parade'' may indicate a potential risk of virus spread while some hot hashtags like ``\#StayHome'' or ``\#GetVaccine'' may represent the safety awareness of individuals in the prevailing areas. 
Over the past decades, researchers have successfully applied social media data to monitor the earthquakes~\cite{sakaki2010earthquake} or air quality~\cite{jiang2019enhancing}. 
Inspired by these works, we aim to incorporate social media content to forecast the pandemic.

To this end, we want to answer the following interesting research questions:
\begin{itemize}
    \item \textit{Can social media contents further enhance the short-term and long-term COVID-19 forecasts?}
    \item \textit{How to identify potential risk factors from the social media data as these factors may vary over time and space?}
\end{itemize}
Motivated by them, we collaborate with Twitter and use their COVID-19 stream API service to crawl large-scale tweets related to COVID-19 based on Twitter's internal COVID-19 annotations.
We propose a novel framework, \modelfullname (\modelname), which is composed of two modules, information extraction module and time series prediction module. Specifically, in the information extraction process, we recognize named entities and identify relationships among them from the large-scale tweet corpora. Based on the entities and relationships, we build a spatial-temporal heterogeneous knowledge graph.
We then propose a Dynamic Graph Neural Network (DGNN) with a Bidirectional Recurrent Neural Network (Bi-RNN) to forecast pandemic trends and suggest risk factors for each location.

Our main contributions are summarized as follows:
\begin{itemize}
    \item To the best of our knowledge, we are the first to simultaneously detect social events for pandemic surveillance and suggest the risk factors.
    \item We propose a novel framework, \modelname, for domain-specific information extraction from social media data and time series prediction on dynamic spatial-temporal graphs. Extensive experiments show the effectiveness of our approach. We achieve 7.3\% and 7.4\%  improvements from the state-of-the-art methods for confirmed case/fatality predictions.
    \item We will open-source our implementations to facilitate the research community.
\end{itemize}

\section{Related Work}

\subsection{Pandemic Forecast}
\noindent\textbf{Epidemic Prediction Models.}
There are three types of epidemic prediction models in literature, including exponential growth models~\cite{livadiotis2020statistical}, self-exiting branching process~\cite{kong2021evently}, and compartment models~\cite{ross1916application,aron1984seasonality,harko2014exact,beckley2013modeling,kroger2020analytical,schlickeiser2021analytical,kermack1927contribution,bailey1975mathematical,mohamed2021new,biswas2014seir,hethcote2000mathematics,fine2011herd}. The dynamics of infectious diseases are expressed by the compartment models for predicting the epidemic trends using ordinary differential equations~\cite{ross1916application}. SIR~\cite{kermack1927contribution}, as the most prevailing compartment model, segments the population into three parts: Susceptible, Infectious, and Recovered and express the population flow among them with evolving equations. Later, many cumulative studies based on SIR emerge, including SEIR~\cite{aron1984seasonality}, SEIS~\cite{wan2007seis}, MSEIR~\cite{hethcote2000mathematics}, SuEIR~\cite{zou2020epidemic}, and MSIR~\cite{mohamed2021new}. In specific, SEIR includes the Exposed compartment and SEIS, MSIR, MSEIR, SuEIR extend SEIR by taking into account either Immunity or untested/unreported compartments. However, as concluded in \cite{bertozzi2020challenges}, the exponential growth models can only address the initial outbreak while self-exiting-branching process and compartment models favor the development and peak stages. None of these models are expected to be precise and robust in the long-term pandemic prediction. 

\noindent\textbf{Statistical and Machine Learning Models.} Researchers also apply statistical time series prediction models such as ARIMA and PROPHET for COVID-19 pandemic prediction~\cite{kufel2020arima,mahmud2020bangladesh}. ARIMA~\cite{box2015time} is an Autoregressive Integrated Moving Average model, relying on a basic assumption that the future time series are linear aggregations of the past ones. PROPHET~\cite{taylor2018forecasting} is an additive model that emphasizes seasonal effects so that the model works better on time series with periodical patterns. 
\citet{saba2020forecasting,rodriguez2020deepcovid,chimmula2020time} aggregate neural networks to an Autoregressive model, to enhance inter-region connections or temporal dependencies.
However, these models conduct pandemic forecasts highly depending on the trend and seasonality instincts behind the historical COVID-19 statistics, incapable of incorporating heterogeneous features. \citet{panagopoulos2020transfer,gao2021stan,jin2021inter} apply graph neural networks to take advantage of the mobility data across different regions but still cannot detect hidden risk factors for the pandemic modeling. Therefore, in this paper, we propose a social media enhanced pandemic forecast framework to incorporate the extracted entities and relationships for confirmed case/fatality prediction with strong interpretability.

\subsection{Prediction with Social Media Data}
Plenty of studies have utilized the social media data for various prediction tasks including air pollution monitoring~\cite{mei2014inferring,hswen2019feasibility,jiang2015using,kay2015can,jiang2019enhancing}, earthquake forecast~\cite{sakaki2010earthquake,wang2019rapid}, stock market prediction~\cite{pagolu2016sentiment,jin2017tracking}, and disease detection~\cite{sadilek2016deploying,mcclellan2017using,chen2017forecasting}. However, limited work incorporates the social media data to calibrate the COVID-19 pandemic surveillance. \citet{qin2020prediction} employ the search index of Baidu search engine to serve as a pandemic early predictor. \citet{bae2021accounting} leverage the social effect of media information to strengthen the compartment model for pandemic prediction. However, this study solely takes into consideration the social effects of the media to users' normal life while our method curate every tweet and detect significant social events to enhance the pandemic prediction. 
\begin{figure*}[t]
    \centering
    \includegraphics[width=\linewidth]{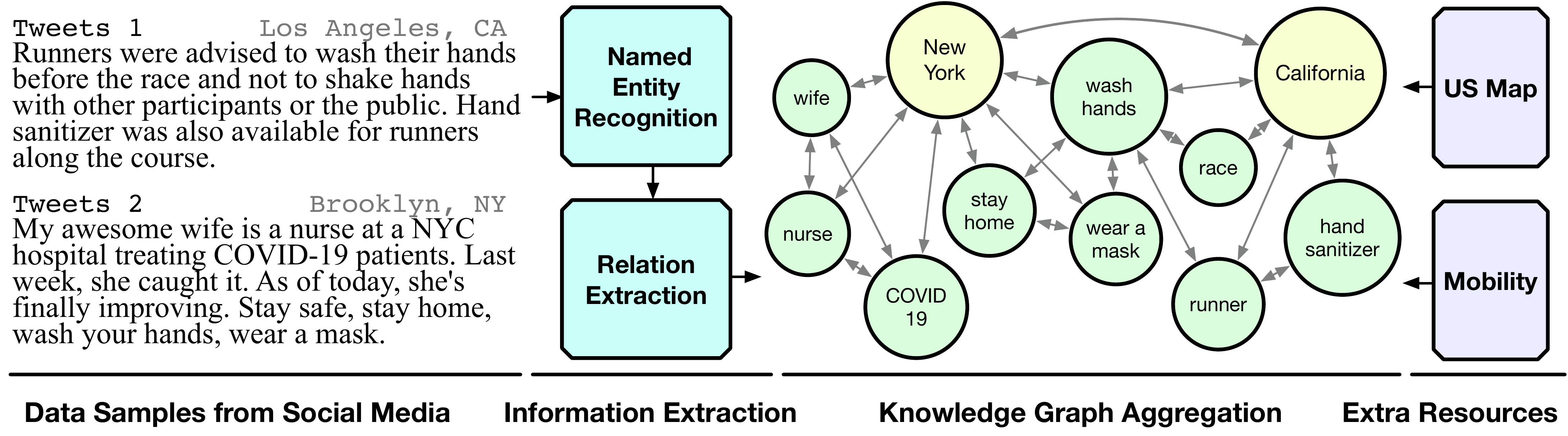}
    \caption{Overview of the information extraction pipeline on social media data. }
    \label{fig:ie}
\end{figure*}

\section{Social Media Enhanced Pandemic Surveillance}
Given a large-scale collection of social media data together with the historical confirmed cases/fatalities and the population mobility statistics, we aim to forecast the pandemic trend and recognize potential risk factors. 
The framework of our \modelname model consists of two components: (i) information extraction module including a named entity recognizer and a relation identifier (as shown in Figure~\ref{fig:ie}); (ii) spatial-temporal dynamic graph encoder for pandemic trend forecast (as shown in Figure~\ref{fig:dgat}). 

\subsection{Constructing Dynamic Knowledge Graphs from Social Media Data}
We propose a bottom-up solution to extract entities and relations to construct the heterogeneous dynamic knowledge graphs. 

\noindent\textbf{Named Entity Recognition (NER).} 
NER is a natural language processing (NLP) task which labels the tokens in a sequence with tags from a desired tag pool. In this work, we adopt the NER setting to extract entities of interest from the social media data by labeling the words or phrases in the tweet sentences. 
As examples in Figure~\ref{fig:ie}, we want to recognize \textit{nurse} as OCCUPATION, \textit{stay home} as INDIVIDUAL\_BEHAVIOR, \textit{race} as EVENT, and so on. 

Traditional NER approaches~\cite{carreras2002named,florian2003named,passos2014lexicon} heavily rely on expensive and time-consuming feature engineering including parsing the Part-of-Speech tags of each word and the syntactic dependency structures of the sentences. Some recent studies~\cite{collobert2011natural,huang2015bidirectional,liu2018empower} incorporate neural networks with statistical models,  such as conditional random fields~\cite{lafferty2001conditional}, to improve the model performance. With deep language models like BERT~\cite{devlin2018bert} and RoBERTa~\cite{liu2019roberta}, the NER performance can be further improved. Without the loss of generality, we leverage BERT model to provide contextualized embeddings and learn a supervised named entity recognizer. To overcome the problem with the nonexistence of annotated tweets as training data, we collect the benchmark corpora and their annotations for multiple NER tasks, including I2B2-2010~\cite{de2011machine}, CORD-NER~\cite{wang2020comprehensive} and MACCROBAT-2018~\cite{caufield2019comprehensive}.
Based on those external datasets, we jointly learn a recognition model to extract entities on the COVID-19 related tweets data. 
On average, we extracted 10,040 unique entities of 45 entity types from 270k tweets corpus every day. 

\noindent\textbf{Relation Extraction.}
Given the extracted entities, the next step is to identify the relationships among the entities. Note that we only extract intra-tweet relations.
In other words, we do not predict the relation between entities in different tweets.
Existing solutions~\cite{verga2018simultaneously,lever2017painless,panyam2018exploiting,zhang2019exploring} formulate the problem as a sequence classification task, given a textual sequence and the positions of two named entities. Specifically, a multi-class classification is conducted to assign a label from a desired set for the relationship. However, this formulation highly depends on the quality and quantity of the annotated datasets to achieve satisfactory performance. It is obviously incapable of identifying emerging new relation types.

To overcome the above challenge, we convert the multi-class prediction task to a binary classification problem of only identifying the existence of a potential relationship between any entity pair in each tweet instance. We aggregate datasets from multiple tasks including Wiki80~\cite{han2019opennre}, I2B2-2012~\cite{sun2013evaluating}, and MAACROBAT-2018~\cite{caufield2019comprehensive} to create the positive training data (labeled as `True'). In order to achieve balanced training, validation and test datasets, we apply negative sampling to create the same number of instances with the label `False'. Note that we assume no relation between any two entities exists if the entities were not annotated. Similarly, we acquire the sequence representations from the fine-tuned BERT language model and feed them into a binary classification layer for label prediction. During the inference stage, we enumerate all possible pairs of entities in each tweet and assign binary labels for them.

\noindent\textbf{Domain-specific Pre-trained Language Model.}
To tackle domain-specific tasks, such as Clinical information extraction~\cite{zhou2021create} and Bioinformatics knowledge acquisition~\cite{lan2018survey}, recent studies pre-train new language models with large-scale corpora collected from those domains~\cite{lee2020biobert,alsentzer2019publicly} to learn customized token and sequence representations. Motivated by these approaches, we leverage all COVID-19 relevant text corpora together with the social media data to pre-train a \texttt{CoronaBERT} language model with 12 layers of Transformers and over 110 million parameters, in order to equip our models with powerful input embeddings. We ceaselessly fine-tune the parameters in \texttt{CoronaBERT} as more COVID-19 stream corpora become available and release the models on a quarterly basis. 

\noindent\textbf{Heterogeneous Knowledge Graph Aggregation.}
After named entity recognition and relation extraction, we apply the DBSCAN clustering model~\cite{ester1996density} to merge semantically similar entities for reducing the noises in the entity sets. This step is essential for cleaning the entities extracted from tweets. For example, ``Marathon'' and ``Marathon:)'' %
are supposed to be merged and ``COVID-19'' is indeed the same as ``COVID2019''. In specific, we cluster the entities based on the similarity among their entity embeddings acquired by \texttt{CoronaBERT}. We assign the node in each cluster with the highest occurrence in tweets as the cluster head. Other nodes in the same cluster will be replaced by the cluster head.

Based on the clustering results, we aggregate the denoised knowledge pieces into a heterogeneous knowledge graph. Two types of nodes exist in the graph, including location nodes and entity nodes. Here we set the location nodes as the 50 states in the United States while our methods can be easily extended to the county-level locations or applied to other countries and regions. Next, we build three types of edges as follows: 
\begin{itemize}
    \item \textbf{Entity-Entity} edges: we add an edge between any two entities if there is a `True' relationship identified. 
    \item \textbf{Location-Entity} edges: we look up the geo-location attribute of the tweet where each entity is extracted and add an edge between the entity node and the geo-location.
    \item \textbf{Location-Location} edges: we add an edge between a  location pair under two circumstances, (i) two locations are adjacent to each other on the US map; (ii) we detected population transition from one location to another according to the mobility data. More details of the mobility data are provided in Section~\ref{sec:data}.
\end{itemize}
We build one knowledge graph for each day. Later, knowledge graphs within a certain time period will be further aggregated for time series prediction, as described in Section~\ref{sec:tsp}.

\subsection{Time Series Prediction with Dynamic Graph Attention Network}\label{sec:tsp}

\noindent\textbf{Dynamic graph aggregation.}
We represent the heterogeneous knowledge graph of the $t$-th day as $G^{(t)}=(V^{(t)},E^{(t)})$ where $n=|V^{(t)}|$ denotes the number of nodes, $V^{(t)} = V^{(t)}_L \cup V^{(t)}_E$, where $V^{(t)}_L$ is the location node set and $V^{(t)}_E$ is the entity node set. Given a sequence of knowledge graphs $\{G^{(1)}, G^{(2)}, ... , G^{(T)}\}$ of length $T$, we aim to predict the COVID-19 courses including confirmed cases and fatality cases on the day $T+l$. We regard it as a short-term prediction when $l<14$ or a long-term prediction when $l\geq14$. We formulate the time series prediction problem as a regression task.

\begin{figure}[t]
    \centering
    \includegraphics[width=\linewidth]{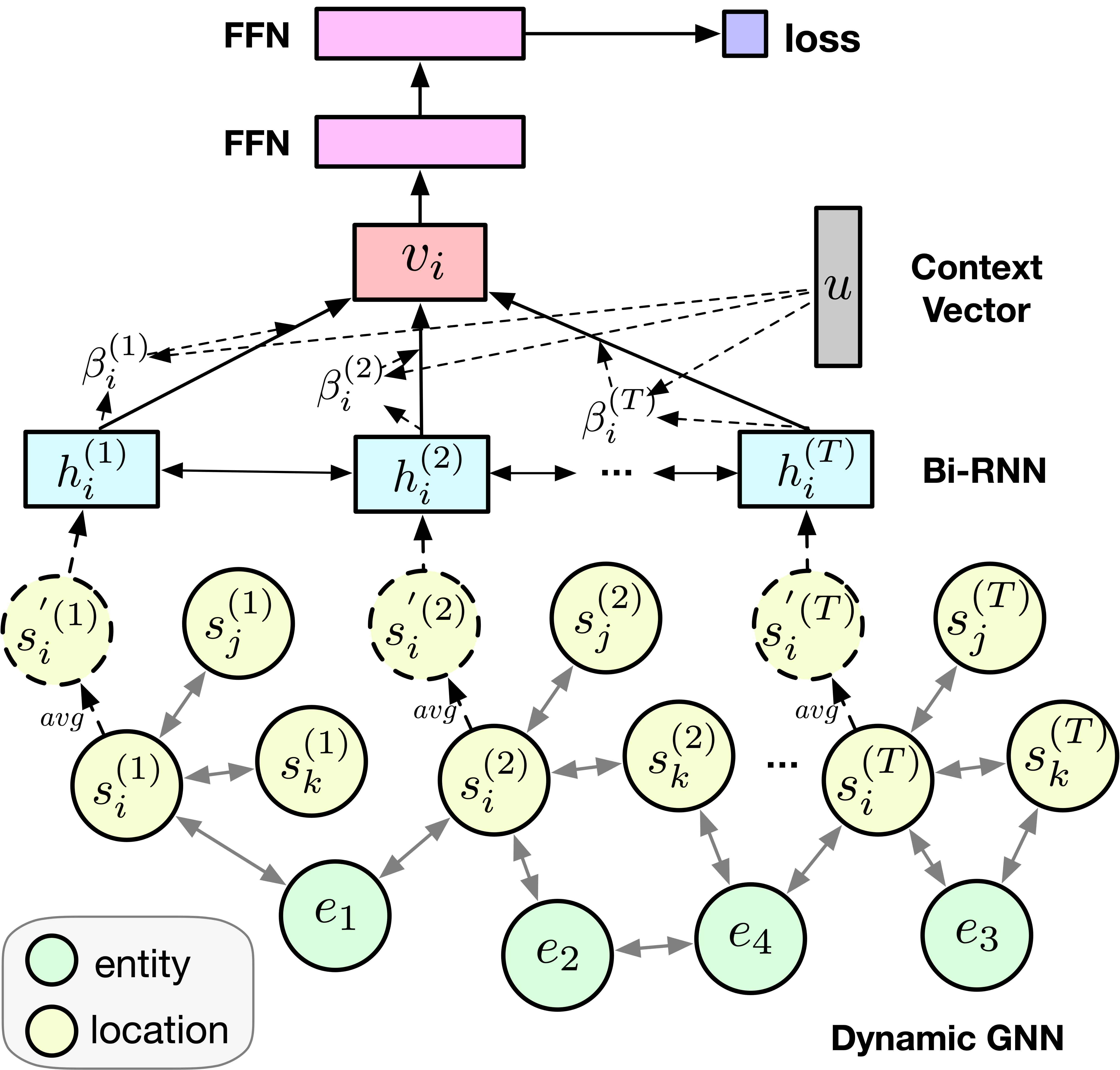}
    \caption{Overview of the time series prediction module.}
    \label{fig:dgat}
\end{figure}

We continue to aggregate the length-$T$ graph sequence into one spatial-temporal graph $G^S=(V^S, E^S)$ as shown in Figure~\ref{fig:dgat}. First, we keep all the location nodes from different times in the period, i.e. $V_L^S = V_L^{(1)} \cup V_L^{(2)} \cup ... \cup V_L^{(T)}$. On the other hand, we merge entity nodes of different times, i.e. $V_E^S = V_E^{(1)} \cup _{\backslash t} V_E^{(2)} \cup _{\backslash t} ... \cup _{\backslash t} V_E^{(T)}$, where $\cup _{\backslash t}$ denotes a time-unaware set union. For example, the entity node $e_1$ is recognized in the location $s_i$ on both time $1$ and time $2$, but we only keep one $e_1$ in $V_E^S$ by connecting $e_1$ to $s_i^{(1)}$ and  $s_i^{(2)}$. In this way, we introduce the inter-time propagation edges to expand the node neighbors along the temporal dimension so that we can easily model the structural temporal dependencies among the nodes. 

\noindent\textbf{Node Features.}
Our pre-trained \texttt{CoronaBERT} is applied to generate the initial semantic features $x^{se}_i$ of dimension $d_{e}$ for node $i$ of any type. We also incorporate the historical COVID-19 statistics $x_{st}$ of $d_{t}$ days ahead of the current time as an extra feature set for location nodes, resulting in a node feature embedding $x_i=x_i^{se} || x_i^{st} $ of dimension $d_{e}+d_{t}$, where $||$ denotes a vector concatenation. Note that we keep the embedding dimensions of location nodes and entity nodes the same, in order to smooth the graph propagation computation. Hence, we append a zero vector of length $d_t$ at the end of each entity vector.

\noindent\textbf{Dynamic Graph Neural Network.} 
We propose a multi-head DGNN architecture to perform the graph propagation. We first conduct a linear transformation on the input node embeddings:
$$z_{i,p} = W_p x_i,$$\\
where $W_p$ is a learnable weight matrix; $p=\{1,...,H\}$; $H$ is the number of heads. Then, we compute a pair-wise un-normalized attention score of an edge between any two neighbors (two nodes $i$ and $j$) in the graph:
$$e_{ij,p} = \text{LeakyReLU}(w_p^T(z_{i,p} || z_{j,p})),$$\\
where $w_p$ is a learnable weight vector and LeakyReLU~\cite{xu2015empirical} is applied as a non-linear transformation. We use the attention score to indicate the importance of a neighbor node in the message passing process, especially when we interpret the risk entities to each location. A Softmax is applied to normalize the attention weights to a probability distribution so that we can easily interpret and compare the importance of all incoming edges, 
$$\alpha_{ij,p} = \frac{\text{exp}(e_{ij,p})}{\sum_{k\in\mathcal{N}_S(i) \cup \mathcal{N}_E(i)}\text{exp}(e_{ik,p})},$$
where $\mathcal{N}_S(\cdot)$ and $\mathcal{N}_E(\cdot)$ denote the sets of neighboring location nodes and entity nodes. We finally aggregate the embeddings of neighboring nodes. The aggregation is scaled by the normalized attention scores. We compute the averaged embeddings over different heads,
$$x'_i = \sigma\left(\frac{1}{H}\sum_{p=1}^H\sum_{j\in\mathcal{N}_S(i) \cup \mathcal{N}_E(i)}\alpha_{ij,p}z_{j,p}\right).$$

\noindent\textbf{Attentive Bi-Recurrent Neural Network.} We intend to further encode the temporal dependencies between the \textbf{location nodes} over times and learn a hidden state of the overall graph using an Attentive Bi-RNN module. We collect embeddings from the same location of different times $[x_i^{'(1)}, x_i^{'(2)}, x_i^{'(T)}]$ and recursively feed them into a Bi-RNN with Gated Recurrent Units (GRU)~\cite{cho2014learning}. We choose GRU instead of Long Short Term Memory (LSTM)~\cite{hochreiter1997long} unit due to its computational efficiency and capability of tackling shorter sequences like tweets~\cite{chung2014empirical}. The hidden representation of each location in time $t$ is learned from two directions,
$$\overleftarrow{h}_i^{(t)} = \text{GRU}(\overleftarrow{h}_i^{(t+1)}, x_i^{'(t)}), \overrightarrow{h}_i^{(t)} = \text{GRU}(\overrightarrow{h}_i^{(t-1)}, x_i^{'(t)}),$$
$$h_i^{(t)}=\overleftarrow{h}_i^{(t)}\oplus\overrightarrow{h}_i^{(t)},$$
We then aggregate the hidden states with another attention mechanism,
$$v_i = \sum_{t=1}^T\beta_i^{(t)}h_i^{(t)}, \beta_i^{(t)}=\frac{\text{exp}(u^Th_i^{(t)})}{\sum_{k}\text{exp}(u^T h_i^{(k)})},$$
where $u$ denotes a context vector and $\beta_i^{(t)}$ are attention scores reflecting the contribution of the hidden representation in time $t$. 

\noindent\textbf{Learning Objective.}
We feed the context-aware node representation $v_i$ into two layers of Feed Forward Networks (FFN) and lastly generate a scalar $\hat{y}_i^{(\bar{t}+l)}$ representing the predicted COVID-19 confirmed case or fatality number in $l$ days ahead of time $\bar{t}$. We compute the loss with the following Mean-Squared-Error (MSE) objective~\cite{ref1},
$$\mathcal{L}=\frac{1}{mn}\sum_{\bar{t}=1}^{n}\sum_{i=1}^{m} (y_i^{(\bar{t}+l)}-\hat{y}_i^{(\bar{t}+l)})^2,$$
where $m$ is the number of location nodes and $n$ is the number of days that requires a prediction.


\section{Experiments}

\subsection{Datasets}\label{sec:data}

\begin{figure}[t]
    \centering
    \includegraphics[width=.95\linewidth]{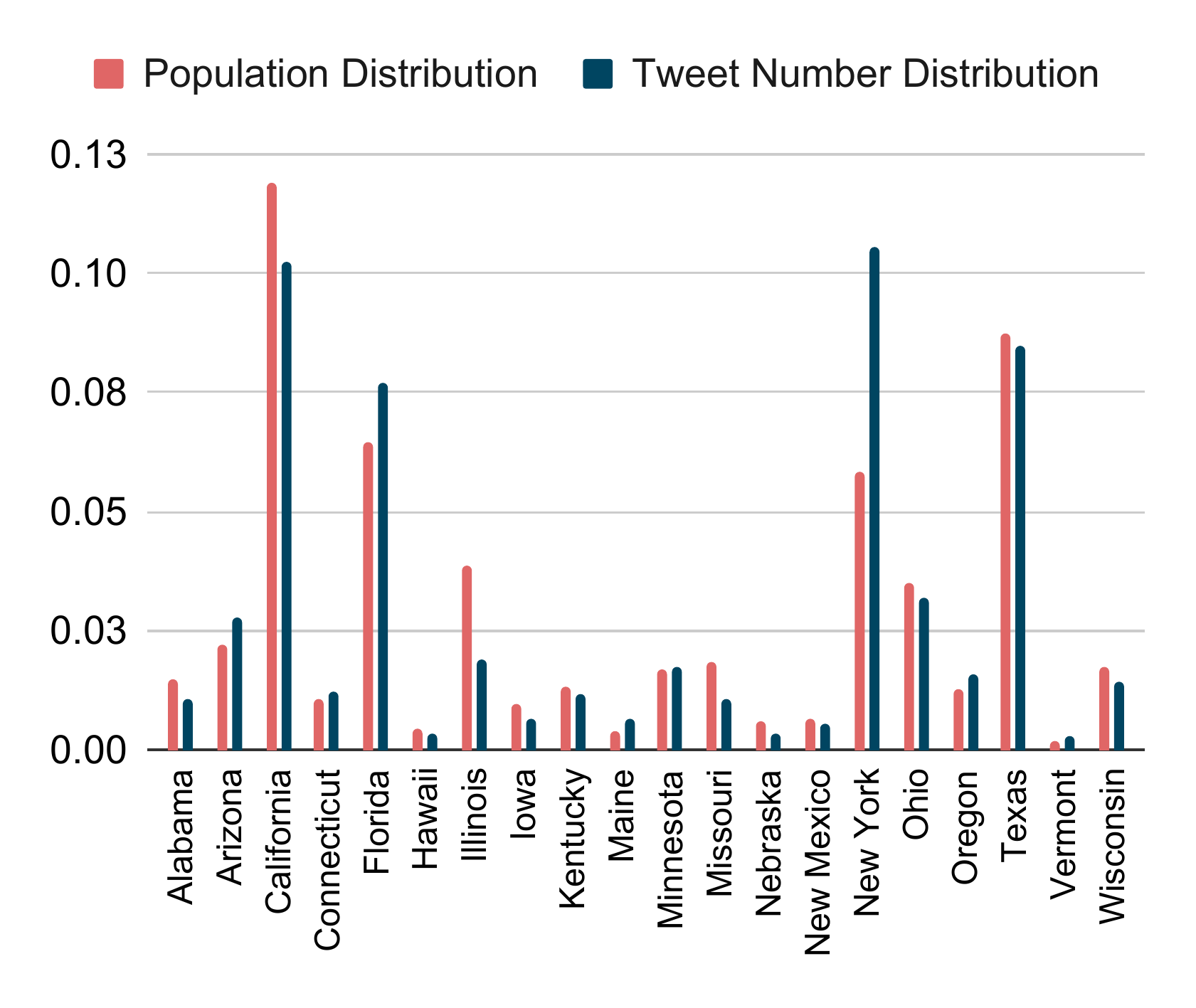}
    \caption{Comparison between the spatial distributions of US population and the number of tweets over 20 states. Each bar represent the percentage of population or tweets in the corresponding state.}
    \label{fig:pop}
\end{figure}

\noindent\textbf{Twitter Stream Data.} 
We collaborate with Twitter and build a real-time tweet crawler to steadily acquire relevant social media tweets using their COVID-19 streaming API\footnote{\url{https://developer.twitter.com/en/docs/labs/covid19-stream/api-reference/}.}~\cite{makice2009twitter}. In detail, the streaming API returns real-time tweets related to COVID-19 based on Twitter's internal COVID-19 tweet annotation system. The data collected for this paper start on May 15, 2020 and end on April 8, 2021. 
Figure~\ref{fig:pop} compares the distributions of the US population and the number of tweets over 20 states. We notice except that New York people are more passionate about posting COVID-19 related tweets while California people do the opposite, other states have relatively similar spatial distributions over the population and number of tweets. 

\begin{figure}[t]
\vspace{-10pt}
    \centering
    \includegraphics[width=.9\linewidth]{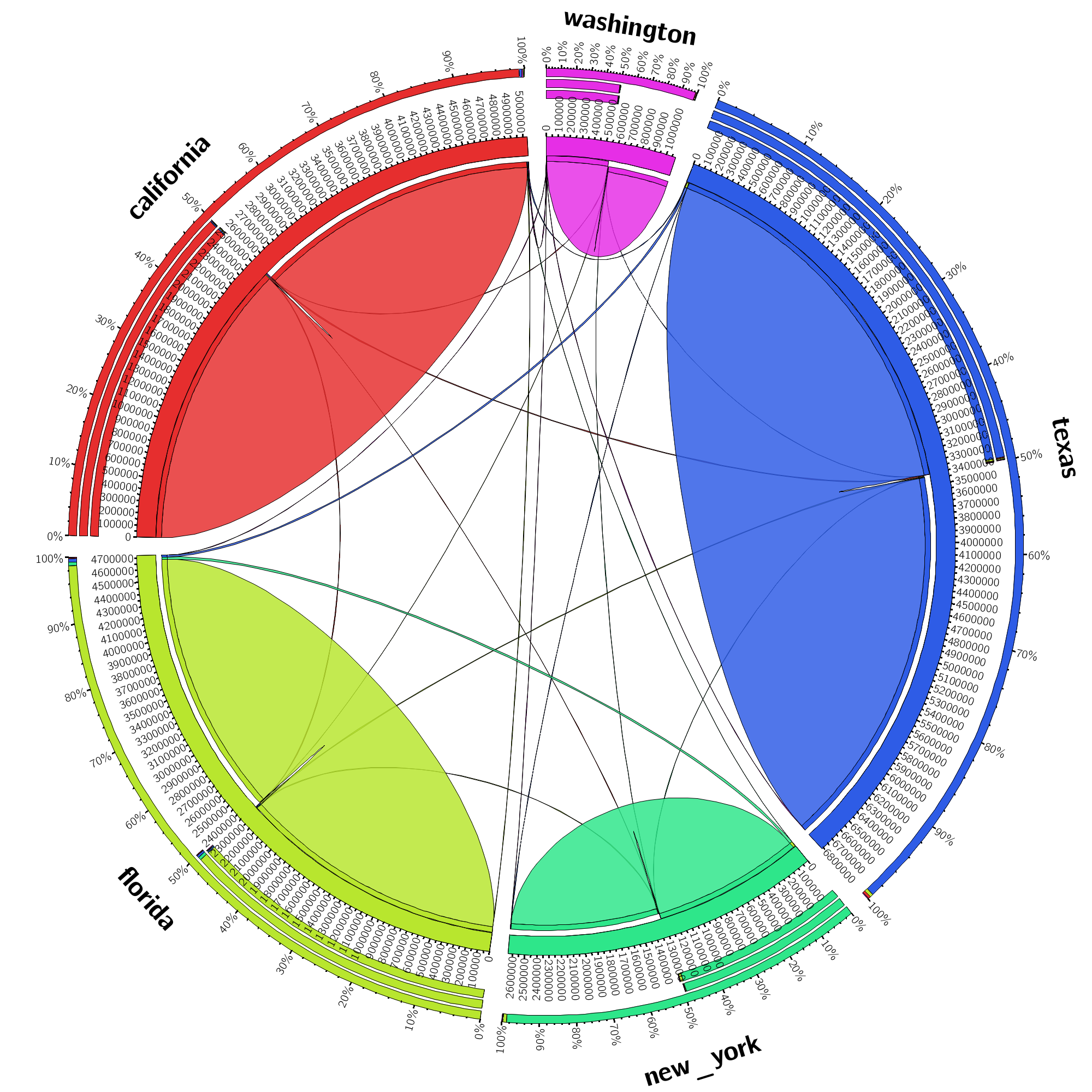}
    \caption{Illustration of a mobility data sample of 5 states on 01-01-2021. Compared to the inter-state transition (black curves), intra-state transition takes the majority (color blocks).}
    \label{fig:circos}
\vspace{-10pt}
\end{figure}


\noindent\textbf{Mobility Data.} As \citet{panagopoulos2020transfer} conclude a strong relationship between the population transition and regional COVID-19 trends, we also collect the mobility data that describe the population transition in the United States from SafeGraph\footnote{\url{https://www.safegraph.com/}.} for pandemic forecast. As shown in figure~\ref{fig:circos}, we illustrate a mobility data sample which includes the population transition among five states on 01-01-2021. The majority transitions are in-state transitions. 

\noindent\textbf{COVID-19 Statistics.} We leverage the US state-level COVID-19 statistics gathered by the New York Times\footnote{ \url{https://github.com/nytimes/covid-19-data}.} based on reports from state and local health agencies for building the ground truths of pandemic forecasts. We use the statistics of confirmed new cases and fatalities from May 5, 2020 to April 8, 2021. Note that the start date is the earliest date when we have Twitter Stream data available. The average new confirmed cases and fatalities over 50 states are $1788.3$ and $28.7$ per day while the standard deviations are $3374.8$ and $63.5$. California has the highest average number of new confirmed cases ($10988.5$) and fatalities ($173.4$). Vermont has the lowest numbers ($60.0$ new confirmed cases and $0.5$ fatalities). 

\subsection{Experimental Setup and Evaluation Metrics}
Following the experimental setup in \cite{panagopoulos2020transfer}, we train a model with the data from time $1$ to time $\bar{t}$ and use it to predict the numbers on time $\bar{t} + l$\footnote{For example, if we predict the next-day (i.e., $l=1$) case number for date 12-31-2020, we make use of all the data between 5-15-2020 and 12-31-2020 to build the training set.}. We evaluate the model on short-term ($l=\{1,7\}$) and long-term ($l=\{14,28\}$) predictions. Note that we learn a different model to predict the cases for time $\bar{t}+l_i$ and $\bar{t}+l_j$, where $i \neq j$. In the training process, we select $5$ data points from the training set as the validation set to identify the best model.  

We evaluate the performance of our method by computing the Mean-Absolute-Error (MAE)~\cite{ref2},
$$\text{error}_{\text{MAE}}=\frac{1}{mn}\sum_{\bar{t}=1}^{n}\sum_{i=1}^{m} |y_i^{(\bar{t}+l)}-\hat{y}_i^{(\bar{t}+l)}|,$$
where $m$ and $n$ denote the numbers of test instances and location nodes. We also follow \cite{kirbacs2020comparative,abbasimehr2021prediction} to compute the symmetric Mean-Absolute-Percentage-Error (sMAPE) to show the average error rate over times and locations, 
$$\text{error}_{\text{sMAPE}}=\frac{1}{mn}\sum_{\bar{t}=1}^{n}\sum_{i=1}^{m} \frac{|y_i^{(\bar{t}+l)}-\hat{y}_i^{(\bar{t}+l)}|}{|y_i^{(\bar{t}+l)}+\hat{y}_i^{(\bar{t}+l)}|}.$$

\subsection{Baselines}

We select three types of baselines and benchmark models to compare to our approach. 

\noindent\textbf{Compartment models.} As there are a large number of compartment models proposed in recent days for COVID-19 forecast, we select three of them with the top performance and complete results in the desired time period from the COVID-19 Forecast Hub\footnote{The model descriptions and up-to-date predicted results can be found at \url{https://github.com/reichlab/covid19-forecast-hub}.}: \texttt{JHU\_IDD-CovidSP}~\cite{lemaitre2020scenario}, \texttt{UCLA-SuEIR}~\cite{zou2020epidemic}, and \texttt{RobertWalraven-} \texttt{ESG}~\cite{esg}. In detail, \texttt{JHU\_IDD-CovidSP} proposes a modified SEIR compartment model where the time in the Infected compartment follows an Erlang distribution to produce more realistic infectious periods. \texttt{RobertWalraven-ESG} is a mathematical model that approximates the SEIR method with a particular skewed Gaussian distribution. \texttt{UCLA-SuEIR} extends SEIR by explicitly modeling the untested/unreported compartment. Note that the 1-day-ahead pandemic forecast results are not provided in the COVID-19 Forecast Hub. 

\noindent\textbf{Statistical time series prediction models.} Two commonly used statistical models are compared to our approach: \texttt{ARIMA} and  \texttt{PROPHET}. \texttt{ARIMA}~\cite{kufel2020arima} is an autoregressive moving average model, explaining a given time series based on its past values. \texttt{PROPHET}~\cite{mahmud2020bangladesh} is a time series prediction model\footnote{\url{https://github.com/facebook/prophet}.} where non-linear trends can be fit with seasonality, plus holiday effects.

\noindent\textbf{Neural network-based models.} A simple two-layer LSTM-based neural network (\texttt{LSTM}) is used for COVID-19 pandemic prediction~\cite{chimmula2020time}, taking the sequence of case numbers from the previous week as the input. \texttt{MPNN}~\cite{panagopoulos2020transfer} is a message passing neural network, building graphs to aggregate the historical case numbers from the neighboring locations based on the mobility magnitude. \texttt{MPNN+LSTM}~\cite{panagopoulos2020transfer} takes advantage of both \texttt{MPNN} and \texttt{LSTM} by jointly learning the graph propagation and temporal dependencies over case numbers of different times. 

\subsection{Implementation Details}
\noindent\textbf{Information Extraction}. We train the named entity recognition and relation extraction models both for a maximum of 10 epochs. The models are implemented in PyTorch and we use Adam optimizer~\cite{kingma2014adam} to optimize the model parameters. We randomly select 10\% instances from the training set as the validation set to select the optimal models. 
To avoid the GPU out-of-memory problem, we filter out tweets with more than 40 words (around 0.17\%). In this work, we focus on the information extraction from English tweets so we also remove the tweets if 90\% of the contents are non-English.  

\noindent\textbf{Time Series Prediction.} We train the model for a maximum of 300 epochs. Early stopping occurs after 100 epochs. Similarly, we utilize PyTorch to implement the model and leverage Adam~\cite{kingma2014adam} for parameter optimization. Batch normalization~\cite{ioffe2015batch} and dropout~\cite{srivastava2014dropout} are applied to the outputs of DGNN and FFN layers to avoid over-fitting. It takes around 8 hours to finish the complete training and evaluation cycle with one NVIDIA V100 GPU.
We employ grid search to find the optimal hyperparameters of our model. Detailed hyperparameter values are listed in Table~\ref{tab:hyper}. 
\begin{table}[t]
    \centering
    \resizebox{.9\linewidth}{!}{
    \begin{tabular}{|l|c|}
    \hline
        Hyperparameter & Value \\
        \hline
        \hline
        Learning Rate & 0.001 \\
        Batch Size & 4 \\
        Dropout Ratio & 0.5 \\
        Bi-RNN Hidden State Size & 64 \\
        DGNN Hidden Unit Size & 64 \\
        Graph Sequence Length $T$ & 7\\
        Semantic Feature Dim. $d_e$ & 768 \\
        Historical COVID-19 Statistics Feature Dim. $d_t$ & 7 \\
    \hline
    \end{tabular}}
    \vspace{10pt}
    \caption{Grid-search is used to find the optimal hyperparameters of our model.}
    \vspace{-10pt}
    \label{tab:hyper}
\end{table}

\begin{table*}[t]
    \centering
    \resizebox{\linewidth}{!}{
    \begin{tabular}{|c|c|c|c|c|c|c|c|c|c|c|}
    \hline
    \multirow{2}{*}{Confirmed Case} & \multicolumn{2}{c|}{1 day ahead} & \multicolumn{2}{c|}{7 days ahead} & \multicolumn{2}{c|}{14 days ahead} & \multicolumn{2}{c|}{28 days ahead} & \multicolumn{2}{c|}{Average}\\
    \cline{2-11}
    & \multicolumn{1}{c|}{MAE} & \multicolumn{1}{c|}{sMAPE} & \multicolumn{1}{c|}{MAE} & \multicolumn{1}{c|}{sMAPE} & \multicolumn{1}{c|}{MAE} & \multicolumn{1}{c|}{sMAPE} & \multicolumn{1}{c|}{MAE} & \multicolumn{1}{c|}{sMAPE} & \multicolumn{1}{c|}{MAE} & \multicolumn{1}{c|}{sMAPE}  \\
    \hline
    \hline
    \texttt{JHU\_IDD-CovidSP}&-&-&1123.721&0.387&1253.138&0.409&1534.643&0.452&1303.834&0.416 \\
    \texttt{RobertWalraven-ESG}&-&-&768.433&0.310&978.533&0.369&2472.093&0.466&1406.353&0.382\\
    \texttt{UCLA-SuEIR}&-&-&755.365&0.258&1099.761&0.335&1591.006&0.439&1148.711&0.344\\
    \hline
    \hline
    \texttt{ARIMA}&604.181&0.200&802.977&0.250&961.297&0.286&1300.487&0.364&917.235&0.275\\
    \texttt{PROPHET}&791.066&0.296&991.049&0.697&1341.798&0.810&2019.242&0.518&1285.789&0.581\\
    \hline
    \hline
    \texttt{LSTM}&1262.333&0.393&1248.080&0.381&1235.201&0.357&1204.188&0.347&1237.450&0.369\\
    \texttt{MPNN}&485.520&0.193&567.745&0.213&825.410&0.266&1304.112&0.352&795.697&0.256\\
    \texttt{MPNN+LSTM}&455.677&0.172&523.770&0.209&672.049&\textbf{0.211}&967.123&0.286&654.655&0.220\\
    \modelname &\textbf{430.007}&\textbf{0.163}&\textbf{474.164}&\textbf{0.203}&\textbf{608.984}&0.216&\textbf{913.202}&\textbf{0.279}&\textbf{606.589}&\textbf{0.215}\\
    \hline
    \end{tabular}}
    \vspace{10pt}
    \caption{Performance of the short-term (1 day \& 7 days ahead) and long-term (14 days \& 28 days ahead) new confirmed case number forecast. All the improvements of \modelname over the baseline methods are statistically significant at a 99\% confidence level in paired t-tests. \modelname achieves 5.6\%, 9.5\%, 9.4\%, and 5.6\% lower MAE than the best baseline \texttt{MPNN+LSTM} when forecasting the new confirmed case numbers for 1, 7, 14, 28 days ahead.}
    \label{tab:case}
\end{table*}

\begin{figure*}[ht]
    \centering
    \vspace{-10pt}
    \includegraphics[width=\linewidth]{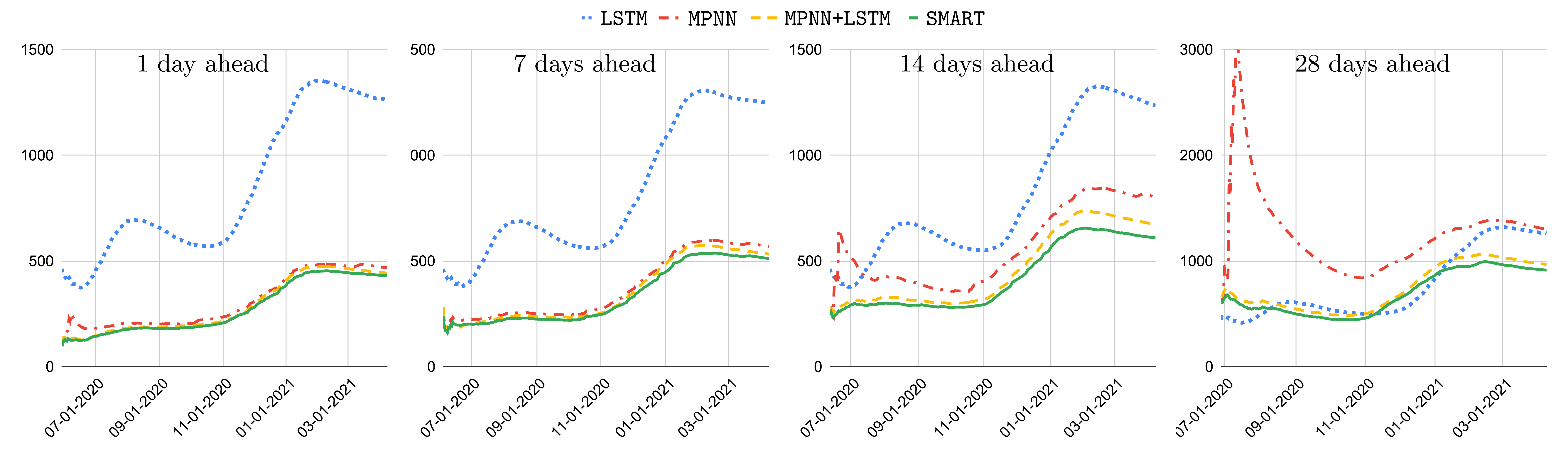}
    \caption{The comparison between \modelname and three neural network-based baselines (\texttt{LSTM}, \texttt{MPNN}, \texttt{MPNN+LSTM}) on the smoothed MAE curve. Each data point on the curve represents the MAE over all the test instances before that date.}
    \label{fig:exp1}
\end{figure*}

\begin{table*}[t]
    \centering
    \resizebox{.95\linewidth}{!}{
    \begin{tabular}{|c|c|c|c|c|c|c|c|c|c|c|}
    \hline
    \multirow{2}{*}{Fatality} & \multicolumn{2}{c|}{1 day ahead} & \multicolumn{2}{c|}{7 days ahead} & \multicolumn{2}{c|}{14 days ahead} & \multicolumn{2}{c|}{28 days ahead} & \multicolumn{2}{c|}{Average}\\
    \cline{2-11}
    & \multicolumn{1}{c|}{MAE} & \multicolumn{1}{c|}{sMAPE} & \multicolumn{1}{c|}{MAE} & \multicolumn{1}{c|}{sMAPE} & \multicolumn{1}{c|}{MAE} & \multicolumn{1}{c|}{sMAPE} & \multicolumn{1}{c|}{MAE} & \multicolumn{1}{c|}{sMAPE} & \multicolumn{1}{c|}{MAE} & \multicolumn{1}{c|}{sMAPE}  \\
    \hline
    \hline
    \texttt{JHU\_IDD-CovidSP}&-&-&18.911&0.465&19.851&0.480&24.362&0.516&21.041&0.487\\
    \texttt{RobertWalraven-ESG}&-&-&15.490&0.452&18.590&0.484&26.179&0.541&20.086&0.492\\
    \texttt{UCLA-SuEIR}&-&-&14.235&0.429&15.603&0.451&19.064&0.495&16.301&0.458\\
    \hline
    \hline
    \texttt{ARIMA} &16.589&0.372&18.649&0.492&22.223&0.437&31.766&0.591&22.307&0.473\\
    \texttt{PROPHET}&19.323&0.423&21.914&0.445&24.469&0.464&29.204&0.500&23.728&0.458\\
    \hline
    \hline
    \texttt{LSTM}&18.039&0.423&17.937&0.432&17.770&0.542&\textbf{17.744}&0.531&17.872&0.482 \\
    \texttt{MPNN}&12.129&0.356&12.897&0.372&14.871&0.380&19.733&0.434&14.908&0.386 \\
    \texttt{MPNN+LSTM}&12.175&0.354&12.785&0.351&14.572&0.379&20.005&0.446&14.884&0.383 \\
    \modelname &\textbf{11.783}&\textbf{0.346}&\textbf{11.847}&\textbf{0.331}&\textbf{13.236}&\textbf{0.349}&18.263&\textbf{0.421}&\textbf{13.782}&\textbf{0.362} \\
    \hline
    \end{tabular}}
    \vspace{10pt}
    \caption{Performance of the short-term (1 day \& 7 days ahead) and long-term (14s day \& 28 days ahead) new fatality number forecast. All the improvements of \modelname over
    the baseline methods are statistically significant at a 99\% confidence level in paired t-tests. \modelname achieves 3.2\%, 7.3\%, 9.1\%, and 8.7\% lower MAE than the best baseline \texttt{MPNN+LSTM} when forecasting the fatality for 1, 7, 14, 28 days ahead.}
    \label{tab:death}
\end{table*}

\subsection{Results}
\noindent\textbf{Confirmed Case Forecast.}
Results of the confirmed case short-term and long-term forecasts are shown in Table~\ref{tab:case}. Compared to the best baseline method \texttt{MPNN+LSTM}, our model improves the average MAE and sMAPE by 7.3\% and 2.3\%, respectively.  
The results show \modelname significantly outperforms the compartment models, such as \texttt{JHU\_IDD-CovidSP} and \texttt{UCLA-SuEIR}. We think the big gap between our method and the compartment models results from the serious over-fitting issue in the SEIR model and its extensions. The SEIR model tends to assume that the peak would come right after the current data and is especially weak at predicting the progression at the early pandemic stage~\cite{gao2021stan}.
We also notice that the two statistical time series prediction models perform differently, and \texttt{ARIMA} gets much lower errors than \texttt{PROPHET} especially in the long-term prediction. This could be because \texttt{PROPHET} is supposed to work best with time series that have strong seasonal effects which is obviously not the situation in the COVID historical statistics. It turns out that a simple linear aggregation over the past case numbers can achieve relatively good performance. 
Besides, \texttt{MPNN} gets higher errors compared to its temporal variant, \texttt{MPNN+LSTM}, denoting the effectiveness of learning the temporal dependencies together with the graph aggregation. However, solely using \texttt{LSTM} to conduct the pandemic forecast achieves quite inaccurate predictions. We think it is because sequence modeling approaches like \texttt{LSTM} are unstable to handle the sequential inputs with sharply changing patterns~\cite{panagopoulos2020transfer}. For instance, it may be hard for \texttt{LSTM} to recognize turning points, such as lockdowns and reopens. 
\modelname initially outperforms other models by a small margin (1-day-ahead forecast) while the improvement increases as the model predicts on later days. 
Compared to \texttt{MPNN+LSTM}, \modelname achieves the largest error reduction of 9.5\% and 9.4\% while forecasting the case numbers in the next 7th and 14th day. 
This could be because the ongoing events discussed on social media would not immediately affect the COVID-19 confirmed case numbers. More precisely, we need 1-2 weeks on average for the newly infected cases to be self-identified, tested and confirmed, based on our observations. 

\begin{figure}[t]
    \centering
    \includegraphics[width=\linewidth]{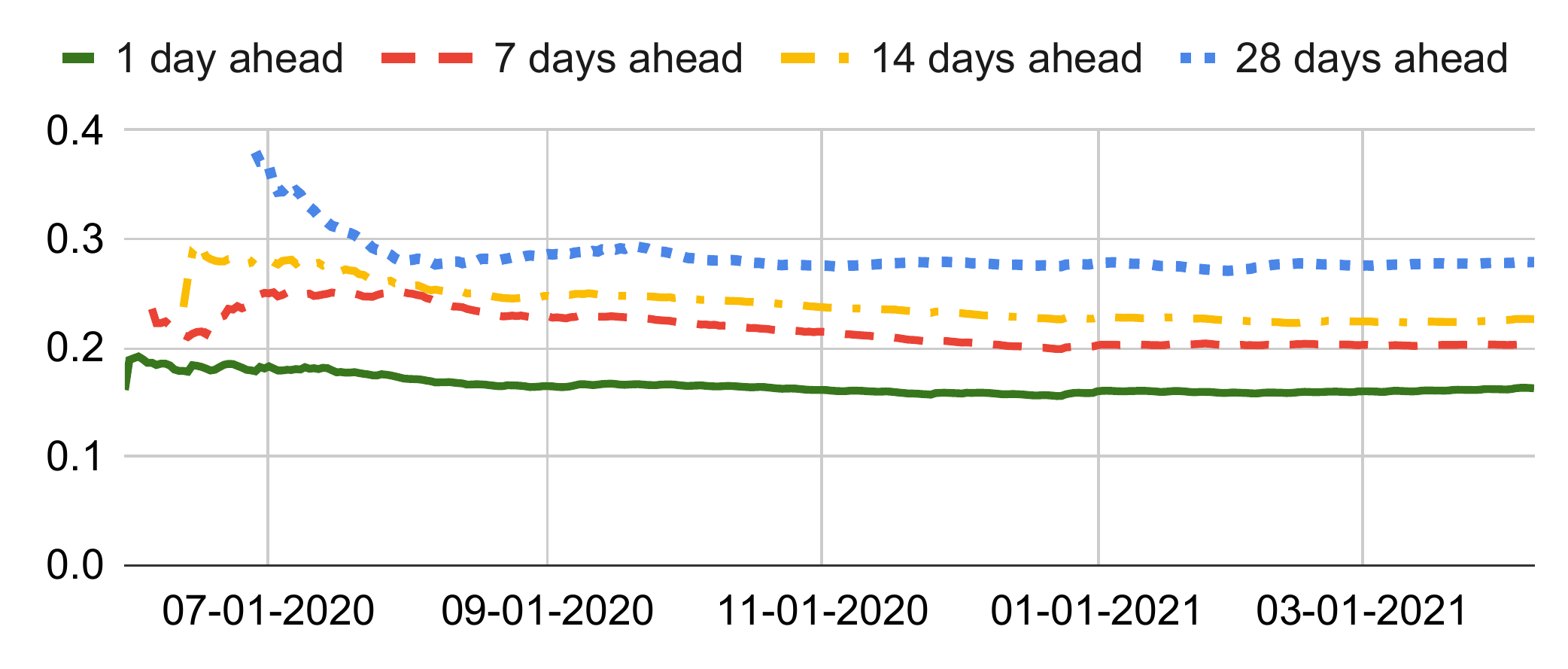}
    \caption{The comparison of smoothed sMAPE curve of \modelname on four forecast tasks. Each data point on the curves represents the sMAPE over all the test instances before that date.}
    \vspace{-10pt}
    \label{fig:smape}
\end{figure}

To observe the detailed forecast performance on every test instance, we plot the smoothed MAE curve for \modelname and three neural network-based baselines (\texttt{LSTM}, \texttt{MPNN}, \texttt{MPNN+LSTM}). Note that every data point on the curves represents the MAE over all the test instances before the corresponding date. We observe that an error explosion becomes more and more clearly visible at the early stage of \texttt{MPNN}. We think \texttt{MPNN} is quite unstable especially when the training data are limited. In contrast, our \modelname model remains stable of all time.
In addition, we observe the average MAE comes to a peak in the middle of January for all the models. This is consistent with the fact that the new confirmed case numbers in the US come to a peak at around the same time. We also plot the smoothed sMAPE curve in Figure~\ref{fig:smape} which shows the sMAPE over the test instances before that date. All the curves quickly converge as the models obtain enough training instances, denoting the stability of our method.

\noindent\textbf{Fatality Forecast.}
We show the results of fatality forecasts in Table\ref{tab:death}. \modelname achieves 7.4\% and 5.5\% lower MAE and sMAPE, compared to the best baseline model \texttt{MPNN+LSTM}.
Among the three compartment models, \texttt{UCLA-SuEIR} performs the best. We surmise that taking unreported/untested cases leads to better modeling on fatalities. 
We notice the MAE of \texttt{LSTM} model is lower than \modelname by 2.9\% while its sMAPE is higher than \modelname by 26.1\%. We believe the \texttt{LSTM} model has been over-fitted to some extremely large or small values so that a large MAE can be avoided but the sMAPE will explode. 
Again, we find that the improvements of \modelname on the 7,14-28-day-ahead forecast tasks (7.3\%, 9.1\%, and 8.7\%) are much more significant than the 1-day-ahead forecast task (3.2\%), demonstrating the long-term advantages of our method.

\subsection{Ablation Study}
We present the ablation study on the 7-day-ahead new confirmed case forecast task to demonstrate the effectiveness of each module in our framework. We observe similar results on other forecast tasks. Here we explain the different settings of our model variants as follows.

\noindent\textbf{w/o RE module.} Under this setting, we exclude the Entity-Entity edges in the heterogeneous knowledge graphs so that we can observe the improvement from our relation extraction module. \\
\noindent\textbf{w/o NER module.} We continue to exclude the Location-Entity edges to check the contribution of our named entity recognition module. Under this setting, all the edge propagation between location nodes and entity nodes are eliminated.\\
\noindent\textbf{w/o Attentive Bi-RNN module.} We remove the Attentive Bi-RNN module from our framework. We alternatively compute an element-wise averaged representation for each location node and feed it into the FNN layer for the pandemic forecast. \\
\noindent\textbf{w/o DGNN module.} To verify the contribution of our DGNN module, we remove the DGNN module but instead recursively feed the sequence of historical COVID-19 statistics features into the Attentive Bi-RNN units for each location node. \\
\noindent\textbf{w/o CoronaBERT Language Model.} We also observe the contribution from our pre-trained CoronaBERT language model by replacing it with a BERT language model (BERT-BASE) to initialize the semantic representations for each node. 

In summary, every component in our framework is proved effective. Removing Entity-Entity and Location-Entity edges leads to 4.3\% and 8.9\% error lifts, respectively. When we jump over the DGNN module, the error dramatically increases, proving the capability of the heterogeneous graph to encode a rich spatial-temporal representation for each location node. The Attentive Bi-RNN module also makes a significant improvement of 10.4\% on the forecast performance.

\begin{table}[t]
    \centering
    \resizebox{\linewidth}{!}{
    \begin{tabular}{|l|r|r|}
    \hline
    Model & MAE & Error Lift(\%)\\
    \hline
    \hline
    \modelname &474.164&-\\
    w/o RE module &495.688&	+4.3\\
    w/o NER module &518.389&	+8.9\\
    w/o Attentive Bi-RNN module &528.025&	+10.4 \\
    w/o DGNN module &1112.334&	+120.8\\
    w/o CoronaBERT Language Model &500.878&	+5.6\\
    \hline
    \end{tabular}}
    \vspace{10pt}
    \caption{Ablation study on the 7-day-ahead forecast task. Similar results can be achieved from other forecast tasks. We can observe significant improvements from all components in our framework.}
    \vspace{-10pt}
    \label{tab:ablation}
\end{table}

\subsection{Risk Factor Discovery}

To identify the potential location-wise \textit{risk factors} of the COVID-19 pandemic, we make use of the normalized attention score $\alpha_{i,j}$ (introduced in Section~\ref{sec:tsp}) which indicates the contribution of each entity node $i$ when node $i$'s message is passed to the location node $j$. For each location, we first rank all the dates based on the number of confirmed cases in decreasing order. We then pick the top 20\% dates with the biggest numbers from all the dates to build a \textit{high set}. 
Ultimately, we aim at discovering a group of significant entities from the tweets that are used to predict the confirmed cases on the dates from the \textit{high set}.
Specifically, during each inference process, we retrieve the attention scores of all the Location-Entity edges for each location node. We then compute a \textit{risk score} for each (Location, Entity) pair by averaging the attention scores over all dates in the \textit{high set}. Finally, the entities with top-$k$ \textit{risk scores} for each location can be considered as the \textit{risk factors}. 

Table~\ref{tab:risk} shows the top-5 \textit{risk factors} of six states: California, New York, Florida, and Ohio, Hawaii, and Vermont with distinct spatial distributions as shown in Figure~\ref{fig:pop}. 
\begin{table}[t]
    \centering
    \resizebox{\linewidth}{!}{
    \begin{tabular}{|l|c|c|c|}
    \hline
    & \textbf{California} & \textbf{New York} & \textbf{Florida} \\
    \hline
    \#1 & pharmacists & traveler & workers\\
    \#2 & \#endthelockdown & doctors & \#stopcovidcorruption\\
    \#3 & mexico city &  test results & crimes\\
    \#4 & covid-positive & bill gates & voting\\
    \#5 & msm & public health & \#stayconnected\\
    \hline
    \hline
    & \textbf{Ohio} & \textbf{Hawaii} & \textbf{Vermont} \\
    \hline
    \#1 & golf & mental health & \#endthelockdown\\
    \#2 & \#hydroxychloroquine & immigrants & rape\\
    \#3 & \#wwg1wgaworldwide & surf & \#wakeupamerica \\
    \#4 & crush & 2ndwave & burger \\
    \#5 & traveler & patients & sickness \\
    \hline
    \end{tabular}}
    \vspace{10pt}
    \caption{Top-5 \textit{risk factors} in six different states related to COVID-19 pandemic.}
    \vspace{-10pt}
    \label{tab:risk}
\end{table}
Some of the entities can be easily connected with the increasing trend of the COVID-19 pandemic. For example, when people are seeking for \textit{ending the lock down} in California and Vermont, or \textit{staying connected} to each other in Florida, they are likely to go out, inevitably facilitating the spread of the virus. When people pay more attention to the local doctor resource or public health condition in New York, the peak of the pandemic should not be far away. However, it may be hard to interpret some entities like \textit{msm} without the contexts since \textit{msm} can be the abbreviation of either \textit{mainstream media} or \textit{master of science in management}. 

We also incorporate the named entity recognition results to show in Table~\ref{tab:type} the top5 \textit{risk factors} under $4$ different categories: HASHTAG, SIGN\_OR\_SYMPTOM, SOCIAL\_INDIVIDUAL\_BEHAVIOR and ORGANIZATION. We notice \textit{msm} is categorized as an organization, so it is more likely to be interpreted as the \textit{mainstream media}. It is obvious that the pandemic is getting more serious if we are facing the \textit{personal protective equipment shortage}. The government and health institutes are better to be prepared if more and more people become sick and have the symptoms such as \textit{cough} and \textit{sneezes}. There are limitations if we only rely on the entities with high attention scores to interpret the \textit{risk factors}. For example, we cannot simply conclude that the prevailing entity \textit{amazon} results in an increasing trend of the pandemic. The relationship between \textit{amazon} and increasing trend might not be causal but just co-occurrence. 
\begin{table}[t]
    \centering
    \resizebox{\linewidth}{!}{
    \begin{tabular}{|l|c|c|}
    \hline
    & \textbf{HASHTAG} & \textbf{SIGN\_OR\_SYMPTOM}  \\
    \hline
    \#1 & \#wakeupamerica & cough\\
    \#2 & \#covidiot & sneezes\\
    \#3 & \#breakingnews &  headaches\\
    \#4 & \#staysafe & chill\\
    \#5 & \#ppeshortage & sickness\\
    \hline
    \hline
    & \textbf{SOCIAL\_INDIVIDUAL\_BEHAVIOR} & \textbf{ORGANIZATION} \\
    \hline
    \#1 & genocide & @youtube \\
    \#2 & loyalty & @nytimes\\
    \#3 & discord & nih \\
    \#4 & voting & amazon \\
    \#5 & racism & msm \\
    \hline
    \end{tabular}}
    \vspace{10pt}
    \caption{Top-5 \textit{risk factors} under four different entity categories related to COVID-19 pandemic.}
    \vspace{-10pt}
    \label{tab:type}
\end{table}

\section{Conclusion and Discussion}
In this paper, we conduct the first trial to incorporate the entities and relationships extracted from social media data to simultaneously enhance the pandemic surveillance and detect the potential risk factors. We propose a dynamic graph neural network to learn the temporal dependency among nodes of different times and propagate the messages among the heterogeneous nodes. Extensive experiments show the effectiveness and robustness of our forecast model. We will open-source our framework and release the pre-trained \texttt{CoronaBERT} language model to facilitate future research in this direction.

Overall, we provide a generic solution for taking advantage of the informative entities and relationships in the social media data. It is straightforward to apply our approach to any future epidemiological surveillance. Our approach also has the potential to tackle other real-world problems, such as environment monitoring and crime detection. In the future, we will focus on detecting the \textit{risk factors} in a more strict manner by identifying the relationship between the \textit{risk factors} and the pandemic trends or predicted targets.  

\begin{acks}
We would like to thank the anonymous reviewers for their helpful comments. This work was partially supported by the National Science Foundation [NSF-DGE-1829071, NSF-IIS-2031187] and the National Institutes of Health [NIH-R35-HL135772, NIH/NIBIB-R01-EB027650].
\end{acks}

\bibliographystyle{ACM-Reference-Format}
\bibliography{ref}

\end{document}